\documentclass{article}
\usepackage{arxiv}

\usepackage{multirow}
\usepackage[table]{xcolor}
\usepackage{array}
\usepackage{tabu}
\usepackage{listings}
\usepackage{multicol}
\usepackage{hyperref}

\usepackage[labelfont=bf]{caption}
\usepackage[labelsep=period]{caption}

\usepackage{enumerate}
\usepackage{amsmath}
\usepackage{graphicx}
\usepackage{subfiles}
\graphicspath{{code/}{../code/}}

\usepackage{caption}

\usepackage{float}
\usepackage[super]{nth}
\usepackage{url}

\usepackage{algorithm} 
\usepackage{algpseudocode}

\usepackage{tikz}
\usetikzlibrary{backgrounds,fit,decorations.pathreplacing, shapes}  
\usepackage{tkz-graph}
\usepackage[utf8]{inputenc}
\usepackage[font=scriptsize,labelfont=scriptsize]{subfig}
\usepackage{subfig}

\usepackage[super]{nth}
\usepackage{adjustbox}
\usepackage{nameref}

\lstdefinestyle{mystyle}{
    basicstyle=\footnotesize,
    breakatwhitespace=false,         
    breaklines=true,                 
    captionpos=b,                    
    keepspaces=true,                 
    numbers=left,                    
    numbersep=5pt,                  
    showspaces=false,                
    showstringspaces=false,
    showtabs=false,                  
    tabsize=2
}
\tikzstyle{vertex}=[circle,draw=black, fill=white,sloped,minimum size=17pt,inner sep=5pt]

\usepackage{tikz}
\usetikzlibrary{decorations.pathreplacing,calc}

\newcommand{\figref}[1]%
{Figure \ref{#1}%
}

\algnewcommand{\LineComment}[1]{\State \(\triangleright\) #1}

\makeatletter
\def\@copyrightspace{\relax}
\makeatother

\title{Density based Community Detection/Optimization}

\author{Rui Portocarrero Sarmento \\
LIAAD-INESC TEC \\
PRODEI - Faculty of Engineering, University of Porto \\
mail@ruisarmento.com
}

\begin{document}

\maketitle

\begin{abstract}
Modularity-based algorithms used for community detection have been increasing in recent years. Modularity and its application have been generating controversy since some authors argue it is not a metric without disadvantages. It has been shown that algorithms that use modularity to detect communities suffer a resolution limit and, therefore, it is unable to identify small communities in some situations. In this work, we try to apply a density optimization of communities found by the label propagation algorithm and study what happens regarding modularity of optimized results. We introduce a metric we call ADC (Average Density per Community); we use this metric to prove our optimization provides improvements to the community density obtained with benchmark algorithms.
Additionally, we provide evidence this optimization might not alter modularity of resulting communities significantly.
Additionally, by also using the SSC (Strongly Connected Components) concept we developed a community detection algorithm that we also compare with the label propagation algorithm. These comparisons were executed with several test networks and with different network sizes. The results of the optimization algorithm proved to be interesting. Additionally, the results of the community detection algorithm turned out to be similar to the benchmark algorithm we used.\end{abstract}

\keywords{Social Networks \and Community Detection Optimization \and Modularity \and Community Density}

\section{Developments and Benchmark}

Several developments were made to test the hypothesis. An algorithm was developed, and a metric is introduced in the following sections.  

\subsection{Average Density per Community (ADC) measure}

Average Density per Community (ADC) is the measure that is used to compare the algorithm results and is given by the following formula:

$$ADC = \frac{1}{n_C} \sum_{C_{i=1}}^{n} Density(Ci)$$

where $n_C$ is the number of communities identified in the graph, $Density(Ci)$ is the density of each community $Ci$.

\subsection{Optimization Algorithm}

Algorithm 1 provides the sequence of tasks we are doing to test the hypothesis \footnote{Available Code at \url{https://github.com/Sarmentor/Density-based-Community-Detection-Optimization}.}. We start by using the results of a community detection algorithm. Then, we try to discover if the communities can be disbanded in smaller communities. These smaller communities are strongly connected components, i.e., groups of nodes with higher density. Then, if the average community density of the disbanded communities is higher than the original community the disbanding is indeed executed. If not, the community founded by the benchmark algorithm is not disbanded and maintains its original id.

\begin{algorithm}
  \caption{Algorithm Pseudo-Code for Optimization of Community Density}
    \begin{algorithmic}[1]
    \renewcommand{\algorithmicrequire}{\textbf{Input:}}
    \renewcommand{\algorithmicensure}{\textbf{Output:}}
        \Require $Communities\_Data$ \Comment{Node List and their Community}
        \Ensure $Community\_Results$ \Comment{New Community Structure}
        
    \While{not at the end of $Original\_Communities$ list}
        \If{$ncomponents > 1$} \Comment{If community has more than 1 component}
            \State $SCC \leftarrow \Call{strong\_connected\_components\_of\_community}{Community_i}$
            \State $mdc \leftarrow \Call{mean\_density\_of\_components}{SCCs}$
            \If{$mdc > Community\_Density$}
            \For{$SCC_i \in Community$}
                
                    \State $Community\_Results \leftarrow \Call{ComponentNodesFormNewCommunity}{SCC_i}$
            \EndFor
        \Else
                    \State $do\_nothing$
        \EndIf    
        \Else
            \State process $next\_community$
        \EndIf  
    \EndWhile
\end{algorithmic}
\end{algorithm}

\subsection{Community Detection Algorithm}

We developed an algorithm for community detection based on density \footnote{Available Code at \url{https://github.com/Sarmentor/Density-based-Community-Detection-Optimization}.}. Moreover, we start by finding strongly connected components in a graph. Then, we initialize all found components to initial communities. After that, we propagate the communities label by aggregating, for each component, the nearby component that maximizes the clustering coefficient of the resulting sub-graph. Algorithm 2 pseudo-code represents the algorithm.

\begin{algorithm}
  \caption{Algorithm Pseudo-Code for Community Detection}
    \begin{algorithmic}[1]
    \renewcommand{\algorithmicrequire}{\textbf{Input:}}
    \renewcommand{\algorithmicensure}{\textbf{Output:}}
        \Require $Graph\_Data$ \Comment{Edge List}
        \Ensure $Community\_Results$ \Comment{Community Structure}
        
        \State $SCC \leftarrow \Call{strong\_connected\_components\_of\_graph}{Graph\_Data}$
        
    \While{not at the end of $SCC$ list}
        \Comment{While there are Components to Explore}
            \For{each $nearby SCC$}
                \If{$Agg\_Subgraph\_CCoefficient = max\_ccoeffient$}
                    \State $Community\_Results \leftarrow \Call{LabelNodesWithSCCLabel}{Nearby\_SCC\_Nodes}$
                \EndIf
            \EndFor
    \EndWhile
\end{algorithmic}
\end{algorithm}

\subsection{Benchmark Algorithm}

The algorithm we used for community detection was the label propagation algorithm. This algorithm, as its name indicates, does the propagation of community label. 

At initial condition, nodes carry a label that denotes the community they belong — belonging to community changes, based on the labels that the neighboring nodes possess. This change is subject to the maximum number of labels within one degree of the nodes. Every node is initialized with a unique label then the labels diffuse through the network. Consequently, densely connected groups reach a common label quickly. When many such dense (consensus) groups are created throughout the network, they continue to expand outwards until it is possible to do so

Label propagation algorithm has the advantage in its running time, an amount of a priori information needed about the network structure (no parameter is required to be known beforehand). The main disadvantage is that it produces no unique solution, but an aggregate of many solutions. Thus, this algorithm suffers from stability issues, and these issues might be adjusted with our optimization algorithm.

We will try to apply our algorithm to the results obtained with the label propagation algorithm and see the changes in ADC and also the modularity of both outcomes.

\subsection{Modularity function}
\label{Modularity_function}

We used a modularity measure $\mathcal{Q}$ to evaluate the quality of the community structure of a graph. Modularity serves as the objective function during the process of calculating the communities \cite{Newman2004}. This measure, apart from  being the most widely used \cite{DBLP:journals/corr/ChenNS15,6693322}, was considered as the quality measure used in the evaluation of the algorithms. Higher values for the modularity $\mathcal{Q}$ mean better community structures. Therefore, the objective is to find a community assignment for each node in the network such that $\mathcal{Q}$ is maximized using the modularity function defined by
\begin{equation}\label{eq:modularity} \mathcal{Q} = \frac{1}{2m}{\sum_{i,j}^{}} \left [ A_{ij} - \frac{{k_ik_j}}{2m}\right ]\delta (c_i, c_j) \end{equation} $A_{ij}$ represents the weight of the edge between $i$ and $j$, $k_i={\sum}_j A_{ij}$ is the sum of the weights of the edges attached to vertex $i$, $c_i$ is the community to which vertex $i$ is assigned, the $\delta$-function $\delta (u, v)$ is 1 if $u=v$ and 0 otherwise and $m=\frac{1}{2}{\sum}_{ij} A_{ij}$. To calculate the modularity of a specific community, the number of inner edges ($in [n]$) and the total number of edges ($tot[n]$) of a specific node $n$ is used. The modularity of the full network can be calculated using the previous $\mathcal{Q}$ function, by considering all the entries of $in$ and $tot$ for all the nodes.

\section{Case Study}

In this case study, we test our optimization algorithm with several toy networks. We used these networks to test the hypothesis that our algorithm indeed provides improvements in the ADC measure in community detection. We also provide a comparison of the modularity results with and without optimization.

Additionally, we test our community detection algorithm with the same networks. We provide a comparison with the label propagation algorithm.  

\subsection{Data Description}

First, we tried optimization with three toy networks. We will call them toynet1, toynet2, and toynet3. These toy networks are directed graphs of small size. The following figures provide a visualization of these graphs. These toy networks have a different disposition of nodes and have strongly connected components. Thus, they offer a good starting point to apply our optimization algorithm.

\begin{figure}[th]
  \centering
  \includegraphics[scale=0.45]{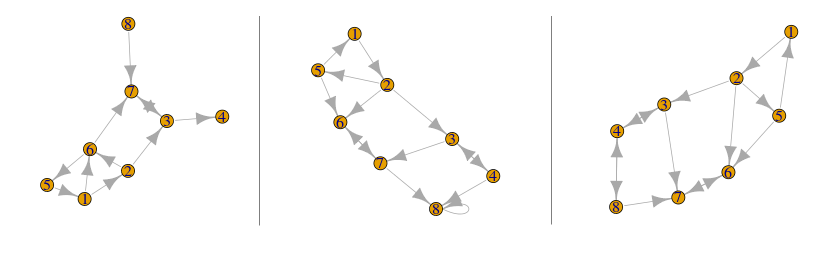}
  \label{fig:toy1}
  \caption{Toy Network 1 (left), Toy Network 2 (middle) and Toy Network 3 (right)}
\end{figure}
  
\subsection{Experiments}

We proceeded by testing both algorithms with several more networks. First, we used three directed toy networks as we previously stated. For the second test, we generated 99 directed networks from the previous toy networks by randomly extracting one edge from each network. This network was a randomly selected network from the set of the three previously described toy networks. This way we would get a mixture of new graphs and also might happen that some graphs are repeated among all the 99 graphs. Thus, we have the possibility of testing also if the algorithms behavior is coherent for different runs.

\subsubsection{Larger Networks}

Additionally, since the generated networks are small size networks we did a third test with more extensive networks. The results of these three tests are presented in the next section.

\section{Results}

This section presents the results of the experiments with our algorithms and the benchmark algorithm. Thus, we show results for ADC and community detection for directed networks. 

\subsection{Community Optimization}

With the toy networks the optimization algorithm had the results represented in the following table:

\begin{table}[h]
\centering
\captionof{table}{Community Optimization}
\begin{tabular}{ |c|c|c|c| }
\hline
\multicolumn{4}{|c|}{Results} \\
\hline
\parbox[c]{1.5cm}{\raggedright Networks} & Original ADC & New ADC & \parbox[c]{1.5cm}{\raggedright Number of new communities} \\
\hline
Toynet1 & 0.238 & 0.708 & 3 \\
Toynet2 & 0.250 & 0.833 & 4 \\
Toynet3 & 0.650 & 0.833 & 3 \\
\hline
\end{tabular}
\label{table1}
\end{table}

\begin{figure}[h]
  \centering
  \includegraphics[scale=0.55]{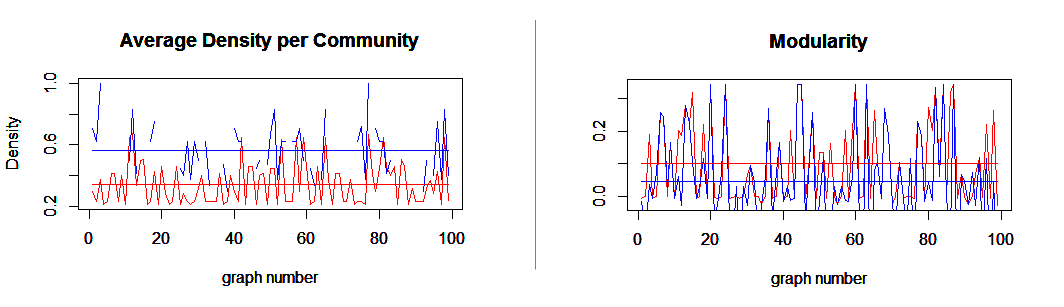}
  \caption{ADC measure (left) and Modularity results (right) for 99 graphs - comparison between ADC optimization algorithm results (blue) and label propagation results (red)}
  \label{fig:adc99}
\end{figure}

For the second test with the 99 networks the following \figref{fig:adc99} exposes the improvements in the ADC measure with our algorithm and comparing with the label propagation algorithm results. 

The blue lines indicate our algorithm results and the red lines show the label propagation results. The horizontal lines provide the average of each graph.

With these results, it is clear that the algorithm provided better ADC than the ADC presented by the original communities provided with the label propagation algorithm. Additionally, regarding modularity, the changes after the optimization are not very significant.

\subsection{Community Detection}

With the toy networks the community detection algorithm had the results represented in the following tables:

\begin{table}[H]
\centering
\captionof{table}{Community Detection - 1st simulation}
\begin{tabular}{ |c|c|c| }
\hline
\multicolumn{3}{|c|}{Modularity Results} \\
\hline
\parbox[c]{1.5cm}{\raggedright Networks} & Label Propagation Algorithm & Our algorithm \\
\hline
Toynet1 & 0.260 & 0.202\\
Toynet2 & 0 & 0.252\\
Toynet3 & 0.119 & 0.253\\
\hline
\end{tabular}
\label{table1}
\end{table}

Then, we did a second, third, fourth and fifth simulation. The second simulation had the following table results:

\begin{table}[th]
\centering
\captionof{table}{Community Detection - 2nd simulation}
\begin{tabular}{ |c|c|c| }
\hline
\multicolumn{3}{|c|}{Modularity Results} \\
\hline
\parbox[c]{1.5cm}{\raggedright Networks} & Label Propagation Algorithm & Our algorithm \\
\hline
Toynet1 & -0.004 & 0.202\\
Toynet2 & 0 & 0.252\\
Toynet3 & 0.283 & 0.253\\
\hline
\end{tabular}
\label{table1}
\end{table}

\begin{table}[th]
\centering
\captionof{table}{Community Detection - 3rd simulation}
\begin{tabular}{ |c|c|c| }
\hline
\multicolumn{3}{|c|}{Modularity Results} \\
\hline
\parbox[c]{1.5cm}{\raggedright Networks} & Label Propagation Algorithm & Our algorithm \\
\hline
Toynet1 & -0.004 & 0.202\\
Toynet2 & 0 & 0.252\\
Toynet3 & 0.283 & 0.253\\
\hline
\end{tabular}
\label{table1}
\end{table}

\begin{table}[th]
\centering
\captionof{table}{Community Detection - 4th simulation}
\begin{tabular}{ |c|c|c| }
\hline
\multicolumn{3}{|c|}{Modularity Results} \\
\hline
\parbox[c]{1.5cm}{\raggedright Networks} & Label Propagation Algorithm & Our algorithm \\
\hline
Toynet1 & -0.004 & 0.202\\
Toynet2 & 0.283 & 0.252\\
Toynet3 & 0.283 & 0.253\\
\hline
\end{tabular}
\label{table1}
\end{table}

\begin{table}[th]
\centering
\captionof{table}{Community Detection - 5th simulation}
\begin{tabular}{ |c|c|c| }
\hline
\multicolumn{3}{|c|}{Modularity Results} \\
\hline
\parbox[c]{1.5cm}{\raggedright Networks} & Label Propagation Algorithm & Our algorithm \\
\hline
Toynet1 & -0.004 & 0.202\\
Toynet2 & 0 & 0.252\\
Toynet3 & 0.017 & 0.253\\
\hline
\end{tabular}
\label{table1}
\end{table}

\begin{figure}[th]
  \centering
  \includegraphics[scale=0.55]{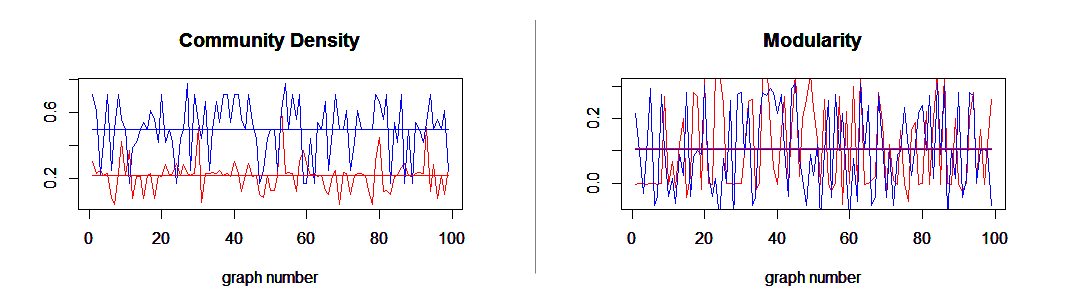}
  \caption{ADC results (left) and Modularity results (right) for 99 graphs - comparison between our community detection algorithm results (blue) and label propagation results (red)}
  \label{fig:mod99}
\end{figure}

After these five simulations, we suspect the label propagation algorithm is very unstable regarding its results. This is clear with the changes in the modularity results for each network simulation. Our algorithm seems to be more stable and presents the same result for every simulation and each tested network. Thus, we proceeded with more tests and with the 99 networks previously generated. The results for modularity are visible in \figref{fig:mod99}.

After previous measurements, it is visible that our algorithm presents similar results on average when compared with the label propagation algorithm. This is true for average modularity.

\subsubsection{ADC Comparison}

We measured ADC for both the label propagation results and the developed algorithm for community detection. The figures provide evidence that the developed algorithm improves ADC measure in large scale without loss of modularity when compared with the benchmark algorithm.

This is an expected result since the basis of our algorithm for community detection is the primary detection of strongly connected components which are inherently regions of high density in social networks.

\section{Conclusions and Future Work}

There is clear evidence our density optimization algorithms provide improvements when the label propagation results are compared with the optimizations results. This is true with our test networks that are directed and small networks. Additionally, robustness tests were done to ensure better that the algorithms are not unstable. These tests proved to be successful, and when we used a more considerable amount of test networks,  we still obtained better ADC results.

Regarding the community detection algorithm, in several tests we performed, it achieved similar average results for the modularity measure which might indicate it is a fairly reasonable algorithm for directed networks. Additionally, this algorithm provides community detection based on dense components, and therefore the density per community (ADC) is higher than the compared benchmark algorithm.

\section*{Acknowledgments}
This work was fully financed by the Faculty of Engineering of the Porto University. Rui Portocarrero Sarmento also gratefully acknowledges funding from FCT (Portuguese Foundation for Science and Technology) through a Ph.D. grant (SFRH/BD/119108/2016). The authors want to thank also to the reviewers for the constructive reviews provided in the development of this publication.

\bibliographystyle{apalike}
\bibliography{arsi}

\begin{thebibliography}{}

\bibitem[Chen et~al., 2013]{6693322}
Chen, M., Nguyen, T., and Szymanski, B. (2013).
\newblock On measuring the quality of a network community structure.
\newblock In {\em Social Computing (SocialCom), 2013 International Conference
  on}, pages 122--127.

\bibitem[Chen et~al., 2015]{DBLP:journals/corr/ChenNS15}
Chen, M., Nguyen, T., and Szymanski, B.~K. (2015).
\newblock A new metric for quality of network community structure.
\newblock {\em CoRR}, abs/1507.04308.

\bibitem[Newman and Girvan, 2004]{Newman2004}
Newman, M. E.~J. and Girvan, M. (2004).
\newblock {Finding and evaluating community structure in networks}.
\newblock {\em Physical Review E}, 69(2):026113+.

\end{thebibliography}

\end{document}